**Influential Papers in Artificial Intelligence and Paediatrics: Assessing RPYS by Experts Review**

Peter Kokol[1], Jernej Završnik[2,3,4, 5], Helena Blažun Vošner[2,3,6]

1. Faculty of Electrical Engineering and Computer Science, University of Maribor, Maribor, Slovenia
2. Community Healthcare Centre Dr. Adolf Drolc Maribor, Maribor, Slovenia
3. Alma Mater Europaea—ECM, Koper, Slovenia
4. Science and Research Center Koper, Koper, Slovenia
5. Faculty of Natural Sciences and Mathematics, University of Maribor, Maribor, Slovenia
6. Faculty of Health and Social Sciences Slovenj Gradec, Slovenj Gradec, Slovenia

**Abstract:** The use of artificial intelligence in paediatrics has vastly increased in last years. Interestingly, no historical bibliometric study analysing the knowledge development in this specific paediatric field has been performed yet, thus our study aimed to close this gap. References Publication Years Spectrography (RPYS), more precisely CitedReferenceExplorer (CRE) software tool was employed to achieve this aim. We identified 28 influential papers and domain experts validation showed that both, the RPYS method and CRE tool performed adequately in in the identification process.

**Key words:** Artificial intelligence, Paediatrics, Bibliometrics, Influential papers, RPYS, CRE

## INTRODUCTION

The use of artificial intelligence (AI) in medicine can be traced back to 1963 when Rand Corporation published its memorandum on AI and brain mechanisms (Maron 1963). The descriptions of some of the first practical applications were published in 1968. In these applications AI was used to support diagnosing (Paycha 1968), and using decision trees in prognosing drug effects in children (Klein et al. 1968). Few years later Shortliffe et al. (Shortliffe et al. 1975) presented the expert system Mycin, developed in 1971, which was able to identify bacteria causing severe blood infections and to recommend antibiotics. Despite the fact that Mycin outperformed members of the Stanford medical school in the accuracy of diagnosis it was never used in practice due to a legal issue -*who do you sue if it gives a wrong diagnosis?* In the same year when Mycin was developed Bellanti & Schlegel (Bellanti and Schlegel 1971) published a study in which they presented a paediatric AI application supporting the diagnosing of immune deficiency diseases. However, the use of AI in medicine become really popular only in 2016, when the IBM AI platform Watson correctly and spectacularly diagnosed and proposed an effective treatment for a 60-year-old woman's rare form of leukaemia (Brusco 2016). Since then more than 12000 papers presenting AI application in paediatrics were published covering themes like use of machine learning in public health, deep learning in image and signal processing, diagnosing, emotion recognition and serious games for health education (Kokol et al. 2021), indicating that AI become a hot research topic also in paediatrics.

Some recent review papers and also a bibliometric based review analysed the use of AI in paediatrics in general (Jamnik 2019; Kokol et al. 2017; Shu et al. 2019) and further ones reviewing the AI use in paediatrics subspecialities like: radiology (Desai et al. 2021; Otjen et al. 2021), prematurity (Reid and Eaton 2019; Scruggs et al. 2020), surgery (Gödeke et al. 2020; Wang et al. 2020) or cardiology (Gaffar et al. 2020; Gearhart et al. 2020). However, to the best of our knowledge, knowledge development in the use of AI in paediatrics based on influential papers has not been analysed, yet. Nevertheless, the importance of studying history and the development of knowledge in medical fields is well known. It enables medical professionals to learn from past experiences and influence the practice in clinical environments in more efficient ways. The aim of our current study is to close this gap.

**METHODS**

Historical roots as important or influential publications in a specific research area (SRA) were first mentioned by Robert K. Merton ( a prominent scientist sometimes referred as the founder of the modern sociology of science) in 1985 (Merton 1985). Since than, historical roots and influential papers identification become a topic of interest in scientometrics and bibliometrics research. Simply, the number of citations that SRA publications received seems to be an obvious measure to identify SRA's historical roots. However, SRA publications might be (frequently) cited also by publications outside the SRA, indicating that such publication could be influential in other SRAs, but not in the SRA in question. Moreover, the SRA's cited publications may not be indexed in bibliographic databases at all. Consequently, a method called References Publication Years Spectrography (RPYS) has been developed to overcome this problem. One of the core software tools implementing RPYS is the CitedReferenceExplorer (CRE; www.crexplorer.net) (Thor et al. 2016, 2018), which we employed in our study. The method has been already successfully used in different medical SRAs (Blažun et al. 2019; Kokol, Blažun, et al. 2021). CRE analyses references' publication years and aggregate the number of cited references over time on a spectrogram. Pronounced peaks indicates the years when historical roots/influential papers were published. In addition to the spectrogram CRE offers some other tools to identify influential publications. In our study we used the CRE tabular output to identify historical roots for the early period and the CRE spectrogram for the recent period. Additionally, CRE indicators, namely N_TOP10, which reveals publications that were among the 10% most cited publications over a longer period and Sleeping beauties (SB) indicators were also used. An SB is a publication that goes unnoticed (sleeps) for a long time and then almost suddenly become highly cited and turns out to be interesting (awakens). SBs frequently presented important innovations and their awakening can be associated with important paradigm shifts in science (Peter Kokol et al. 2020; van Raan 2004). The identified influential publications were reviewed by five domain experts coming from computer and health science fields to informally asses the accuracy of the identification (Kokol, et al. 2021).

To perform the analysis, we formed a corpora of publications related to application of AI in paediatrics harvested from the SCOPUS bibliographic database using the search string (*"machine learning" or "rough sets" or ("decision tree*" and (induction or heuristic)) or "artificial neural network*" or "support vector machines" or "rough sets" or "deep learning" or "intelligent systems" or "artificial intelligence") and (newborn* or toddler* or child* or adolescent*)* and exported their metadata to the CRE. The search was performed on 3[rd] of October 2021.

## RESULTS AND DISCUSSION

The search resulted in 12360 publications citing 977812 references. After removing duplicates 781526 references remained and were analysed with the CRE. The analysis resulted in 28 historical roots/influential papers, which are presented in Figure 1 and Table 1. The oldest cited reference, namely Hobbes Leviathan (Hobbes 1651) was published in 1651. Regarding the content, the references can be divided into three groups:

- Publications concerned with general intelligence, emotions, psychology, behaviour and functioning of brains,
- Publications concerning statistical and mathematical methods, and
- Computer algorithms, tools and languages used for building artificial intelligent systems.

Knowledge development vent trough several stages, namely (1) researching general intelligence, emotions and attitudes, (2) establishing statistical and mathematical foundations, (3) brain and neural networks research, (4) computing machinery and intelligence (5) development of AI algorithms and their applications in paediatrics, and (6) Deep learning applications in paediatrics.

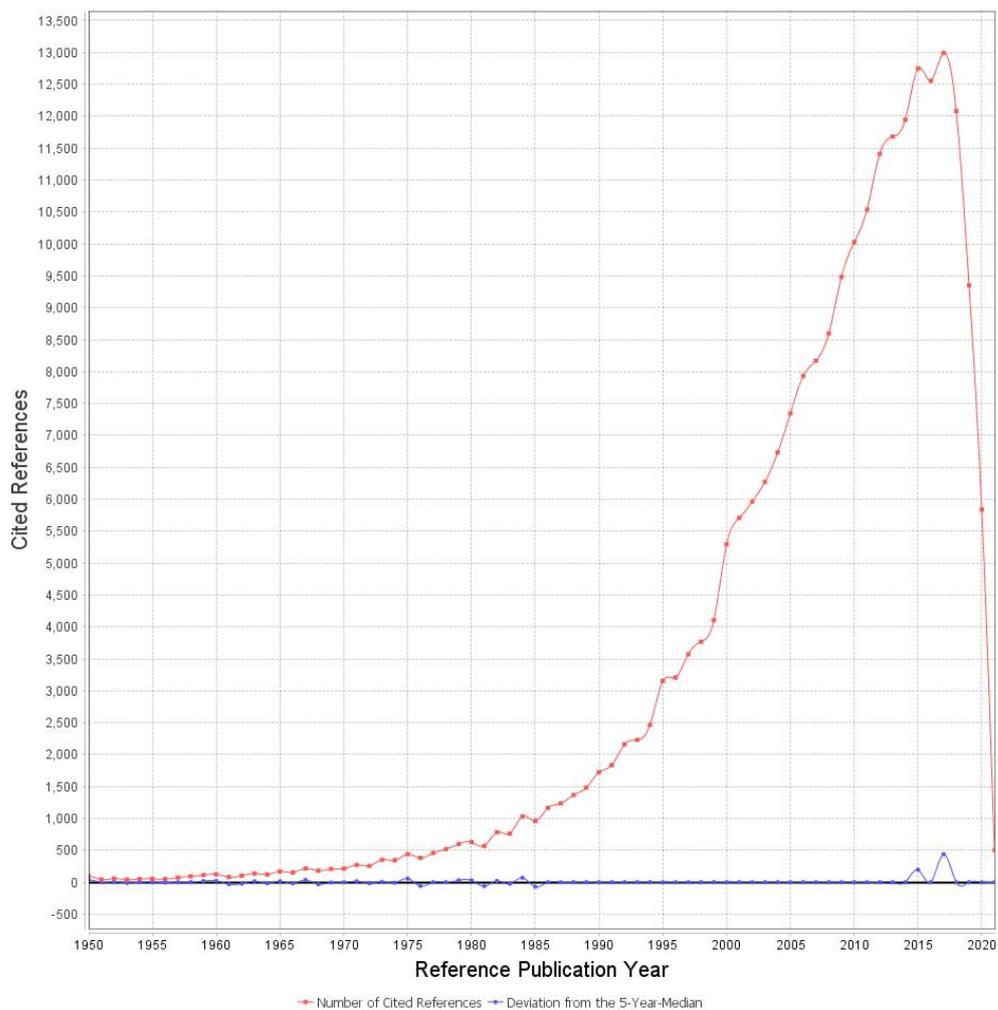

Figure 1. Paediatric historical roots in area of using Artificial intelligence in paediatrics for the period 1950 (the year when A. Turings, Computing machinery and intelligence was published) to 2021. Peaks in blue curve represent years when important publications were published.

Table 1. Historical roots in the area of using Artificial intelligence in paediatrics

| Authors | Year | Title | |
|---|---|---|---|
| Hobbes, T | 1651 | Leviathan | Oldest |
| Darwin, C., | 1872 | The expression of the emotions in man and animals | Tabular |
| James, W., | 1890 | Principles of Psychology | Tabular |
| Pearson, K., | 1901 | On lines and planes of closest fit to systems of points in space | Tabular |
| Spearman, C. | 1904 | General intelligence, "objectively determined and measured" | Tabular |
| Spearman, C. | 1904 | The proof and measurement of association between two things | Tabular |
| Likert, R. | 1932 | A Technique for the Measurement of Attitudes | Tabular |
| Hotelling, H. | 1933 | Analysis of a complex of statistical variables into principal components | Tabular |
| Fisher, R.A | 1936 | The use of multiple measurements in taxonomic problems | Tabular |
| Kanner, L. | 1943 | Autistic disturbances of affective contact | Tabular |
| McCulloch, W.S., Pitts, W | 1943 | A logical calculus of the ideas immanent in nervous activity | Tabular |
| Dice, L.R., | 1945 | Measures of the amount of ecologic association between species | Tabular |
| Hebb, D.O., (1949) | 1949 | The Organization of Behaviour | Tabular |
| Turing, A.M. | 1950 | Computing machinery and intelligence | Spectrogram |
| Zadeh, L.A | 1965 | Fuzzy sets | Spectrogram |
| Cover, T., Hart, P | 1967 | Nearest neighbour pattern classification | Spectrogram |
| Jennett, B., Bond, M | 1975 | Assessment of outcome after severe brain damage | Spectrogram |
| Breiman, L., Friedman, J., Stone, C.J., Olshen, R.A. | 1984 | Classification and Regression Trees | Spectrogram |
| Quinlan, J.R. | 1986 | Induction of decision trees | NTop10 |
| Rumelhart, D.E., Hinton, G.E., Williams, R.J. | 1986 | Learning representations by back-propagating errors | NTop10 |
| Quinlan, J.R. | 1993 | C4.5: Programs for Machine Learning | NTop10 |
| Bishop, C.M | 1995 | Neural Networks for Pattern Recognition | NTop10 |
| Pedregosa, F. | 2011 | Scikit-learn: Machine learning in Python | SB |
| Krizhevsky, A., Sutskever, I., Hinton, G.E. | 2012 | Imagenet classification with deep convolutional neural networks | SB |
| LeCun, Y., Bengio, Y. | 2015 | Deep learning | Spectrogram |
| Esteva, A., Kuprel, B., Novoa, R.A. | 2017 | Dermatologist-level classification of skin cancer with deep neural networks | Spectrogram |

*Assessment*

The expert review revealed that all the identified influential publications might represent works that significantly contributed to the development of Artificial intelligence in general and its use in paediatrics, however some interesting observations were made, which could lead to further investigations:

- Robotic surgery is quite routinely used in medicine, and "emotion teaching" or "stress reduction robots" are widely used in paediatrics, but none influential papers related to robotics or history of robotics were identified. The reason might be that early papers on robotics were not scientific papers, but less known technical reports. Earlier mentions of Greek, Persian or medieval automatons were also not documented written communication (or such documentation was destroyed or lost) and emerged only recently in various histories of AI.

- Statistical and mathematical papers didn't contribute directly to the development of AI, however advanced statistics and mathematics are used in proving of AI algorithms, especially in machine learning.

- There are a lot of influential papers related to prominent machine learning algorithms, like neural networks, deep learning, decision trees and fuzzy sets, and also in paediatrics rarely used algorithms like nearest neighbours, however influenta publications on popular algorithms like support vector machines or Bayes are missing.

## CONCLUSION

Methodologically, RPYS supported by CRE showed to be a feasible methodology to identify historical roots/influential papers. Experts also agreed that the analysis of knowledge development based on historical roots identification can significantly contribute to the understanding of specific research fields, supports learning from important past experiences, provide evidence or motivation for further research and consequently enable health care professionals to influence the everyday practice settings using historical evidence.